\documentclass[prl,showpacs,onecolumn,superscriptaddress,nofootinbib]{revtex4-2}

\usepackage{amsmath,amsthm,amssymb,epsfig,graphicx,mathrsfs,amsfonts,dsfont,bbm}
\usepackage{dcolumn}
\usepackage{bm}
\usepackage{tikz}
\usepackage{graphicx}
\usepackage{xcolor}
\usepackage[colorlinks=true,citecolor=magenta,urlcolor=cyan,linkcolor=black,filecolor=green]{hyperref}
\usepackage[capitalise]{cleveref}
\usepackage{enumitem}
\setlist[itemize]{leftmargin=20pt}
\setlist[enumerate]{leftmargin=20pt}
\usepackage{amsmath,amsfonts,amssymb}
\usepackage{braket}
\usepackage{physics}
\usepackage[linesnumbered,ruled,vlined]{algorithm2e}
\usepackage{algpseudocode}
\usepackage{multirow}
\usepackage{adjustbox}

\theoremstyle{plain}

\theoremstyle{definition}

\newtheorem*{problem*}{Problem}

\newcommand{\vertiii}[1]{{\left\vert\kern-0.25ex\left\vert\kern-0.25ex\left\vert #1
		\right\vert\kern-0.25ex\right\vert\kern-0.25ex\right\vert}}
\newcommand{\Vertiii}[1]{{\vert\kern-0.25ex\vert\kern-0.25ex\vert #1
		\vert\kern-0.25ex\vert\kern-0.25ex\vert}}

\makeatletter
\def\l@subsubsection#1#2{}
\makeatother

\begin{document}

\title{Locality implies Complex Numbers in Quantum Mechanics}
 \author{Tianfeng Feng}
  \affiliation{QICI Quantum Information and Computation Initiative, Department of Computer Science, The University of Hong Kong, Pokfulam Road, Hong Kong}
 \author{Changliang Ren}
\affiliation{Key Laboratory of Low-Dimension Quantum Structures and Quantum Control of Ministry of Education, Department of Physics,
Hunan Normal University,
Changsha 410081,
China}
\author{Vlatko Vedral}
\affiliation{Clarendon Laboratory, University of Oxford, Parks Road, Oxford OX1 3PU, United Kingdom}

\date{\today}
\begin{abstract}
We show that the presented real-number quantum theories, compatible with the independent source assumption, require the inclusion of a nonlocal map. This means that if the independent source assumption holds, in these models, complex-number quantum theory is equivalent to a real-number quantum theory with hidden nonlocal degrees of freedom.
Our results suggest that complex numbers may be indispensable for describing the process involving entanglement between two independent systems. 

\end{abstract}
\maketitle

Although complex numbers are crucial in mathematics, they are not necessary for describing physical experiments, as these experiments are usually expressed in terms of probabilities and thus can be described using real numbers. Fields such as electromagnetism and optics introduce complex numbers for convenience and ease of expression, but their use is not essential. In other words, most physical theories can be completely described using real numbers. However, quantum mechanics is the first theory to be formulated using operators acting on a complex Hilbert space, which has puzzled physicists.

Standard quantum theory utilizes complex Hilbert spaces to represent density matrices, observables, and reversible transformations through linear operators. However, it has long been recognized that there are two hypothetical theories that share many characteristics with standard quantum theory but replace the complex Hilbert spaces with real or quaternionic Hilbert spaces.
In 1936, Birkhoff and von Neumann analyzed the logical structure of quantum theory and pointed out that their hypotheses can be satisfied by real and quaternionic models as well as by the standard complex theory \cite{Birkhoff1936}. Although the complex-number quantum theory has consistently been validated by experimental tests and no experiments have necessitated the use of real theories, researchers continue to seek a more fundamental understanding of the origin of the complex structure, beyond empirical validation.
In 1960, Stueckelberg creatively proposed a rule that allows any quantum system in a complex Hilbert space to be mapped to a real Hilbert space, where the original $d$-dimensional complex space is transformed into a $2d$-dimensional real Hilbert space \cite{Stueckelberg1959,Stueckelberg1960}. This realification of quantum theory has led to a series of developments, including applications in simulating complex quantum systems and self-testing \cite{Myrheim,McKague2009,Rudolph2009,Wootters2013}. Recently, Renou et al. theoretically demonstrated that real quantum theory cannot simulate the experimental results of quantum protocols in network structures \cite{Renou2021}, and two experimental groups have independently verified it \cite{Li2022, Chen2022,Wu2022}. Although these quantum theories with real Hilbert space have been shown not to replace complex quantum theory completely, they are fully compatible with standard quantum theory for individual systems.
It is worth noting that recently, some claims suggest that under the assumption of independent sources (locality), real-number quantum theory may be consistent \cite{Hita,Hoffreumon2025,Volovich}.

Here we briefly review Stueckelberg's rule of quantum theory with real numbers using simple algebraic language. We show that the presented real-number quantum theories, under the assumption of independent sources,  introduce a hidden nonlocal map, which contrasts with locality. 
We further discuss how this equates to complex quantum theory, and demonstrate that its indispensability arises from the phenomenon of entanglement between two independent systems. 

\section{Stueckelberg's rule}\label{s1}

Here we briefly review Stueckelberg's rule of quantum theory with real numbers using simple algebraic language.

Any $d$-dimensional pure state can be expressed as $\ket{\psi} =\sum_{j=0}^{d-1} \psi_j \ket{j}=\sum_{j=0}^{d-1}(a_j+ib_j)\ket{j}$, where $a_j$ and $b_j$ are coefficients of real part and imaginary parf of $\psi_j$. 
For a state vector, one can directly map a $ d$-dimensional quantum state with complex numbers to a $2d$-dimensional quantum state with real numbers, by Stueckelberg's rule \cite{Stueckelberg1960,Myrheim,McKague2009}
\begin{equation}
   \ket{\psi}= \sum_{j=0}^{d-1}(a_j+ib_j)\ket{j} \xrightarrow{R} \tilde{   \ket{\psi}}=\sum_{j=0}^{d-1}a_j\ket{0}\ket{j}+b_j\ket{1}\ket{j}.
\end{equation}
The new real-valued quantum state $\tilde{   \ket{\psi}}$ satisfies the normalization condition since $\sum_j a_j^2+b_j^2=\sum_j \abs{\psi_j}^2$. The bra is defined in a similar way, one has $\tilde{   \bra{\psi}}=\sum_{j=0}^{d-1}a_j\bra{0}\bra{j}-b_j\bra{1}\bra{j}$.
In the following, we donate $      \text{Map}_R(\ket{\psi})=\tilde{\ket{\psi}}, \quad \text{Map}_C(\tilde{\ket{\psi}})=\ket{\psi}.  $

For extending a quantum density matrix to a real-number matrix, one needs to introduce an operator $XZ=\left(\begin{array}{cc}
    0 & -1  \\
    1 & 0
\end{array} \right)$ such that

\begin{equation}
   \rho= \sum_{x,x^\prime} (a_{x,x^\prime}+ib_{x,x^\prime}) \ket{x}\bra{x^\prime} \xrightarrow{R}  \sum_{x,x^\prime}   \frac{1}{2} (a_{x,x^\prime} I +b_{x,x^\prime} XZ) \otimes \ket{x}\bra{x^\prime}   ,
\end{equation}
where $a_{x,x^\prime}$ and $b_{x,x^\prime}$ are the real coefficients of density matrix $\rho$, $X$ and $Z$ are the Pauli matrix. Here we use $\Tilde{\rho}$ to donate the density matrix with real numbers by Stueckelberg's rule. 
Without loss of generally, one can rewrite $\Tilde{\rho}$ as 
\begin{equation}
\Tilde{\rho} = \frac{1}{2}\sum_{x,x^\prime} a_{x,x^\prime}   I \otimes \ket{x}\bra{x^\prime}  + \frac{1}{2}\sum_{x,x^\prime} b_{x,x^\prime}   XZ \otimes \ket{x}\bra{x^\prime} = \frac{1}{2}I\otimes \text{Re}(\rho) + \frac{1}{2}XZ \otimes \text{Im}(\rho).
\end{equation}
Since $\rho$ is Hermitian (it can be represented as a real-number matrix),  $\Tr(\rho)=\Tr(\tilde{\rho}).$
Similarly, for any matrix $A$,  
\begin{equation}
    \Tilde{A}= I \otimes \text{Re}(A) +XZ \otimes \text{Im}(A) = \left(\begin{array}{cc}
       \text{Re}(A)  & -\text{Im}(A) \\
     \text{Im}(A)    & \text{Re}(A)
    \end{array}\right ).
\end{equation} 
For general matrices, we use the following map to denote the relationship between $A$ and $\tilde{A}$,
\begin{equation}
      \text{Map}_R(A)=\tilde{A}, \quad \text{Map}_C(\tilde{A})=A.  
\end{equation}
In the subsequent paper, we will interchangeably use these two equivalent notations. The relationship between $\tilde{\rho}$ and $\tilde{\ket{\psi}}$ is
   $ \tilde{\rho}= \text{Map}_R \left ((\text{Map}_C(\tilde{\ket{\psi}}) \text{Map}_C(\tilde{\bra{\psi}})) \right)$.

\section{Operation in extended real-number matrices of a single system}

In this section, the real-number operations for a single (quantum) system are defined to be consistent with the complex-number operations.

\subsection{Addition and multiplication in extended real-number matrices}

Given any matrix $V_1=A_1+i B_1$ and $V_2=A_2+iB_2$,   the sum $V_2+V_1$ is given as

\begin{equation}
    V_1+V_2=(A_1+iB_2)+(A_2+iB_2)=A_1+A_2+i(B_1+B_2).
\end{equation}
For their extended real matrix $\tilde{V_1}$ and $\tilde{V_2}$, we have 
\begin{equation}
   \tilde{V_1}+\tilde{V_2}=  \left(\begin{array}{cc}
        A_1 &  -B_1\\
        B_1  & A_1
\end{array}\right)+\left(\begin{array}{cc}
        A_2 &  -B_2\\
        B_2  & A_2
    \end{array}\right)=\left(\begin{array}{cc}
       A_1+ A_2 &  -B_1-B_2\\
        B_1+ B_2  & A_1+A_2
    \end{array}\right).
\end{equation}
Clearly, $\tilde{(V_1+V_2)}=\tilde{V_1}+\tilde{V_2}$. That is $V_1+V_2 \rightarrow \tilde{{(V_1+V_2)}}=\tilde{V_1}+\tilde{V_2}$. This indicates that for any matrix addition, the sum of two extended real-number matrices is equal to the extended real-number matrix obtained by adding the two original matrices.

Similarly,  for multiplication, $V_2V_1$ is given as

\begin{equation}
    V_2V_1=(A_2+iB_2)(A_1+iB_1)=A_2A_1-B_2B_1+i(B_2A_1+A_2B_1).
\end{equation}
It is mapped to the real matrix and the multiplication is as follows

\begin{equation}
    \tilde{V_2} \tilde{V_1}=\left(\begin{array}{cc}
        A_2 &  -B_2\\
        B_2  & A_2
    \end{array}\right)
    \left(\begin{array}{cc}
        A_1 &  -B_1\\
        B_1  & A_1
    \end{array}\right)=\left(\begin{array}{cc}
        A_2A_1-B_2B_1 &  -A_2B_1-B_2A_1\\
        A_2B_1+B_2A_1  & A_2A_1-B_2B_1
    \end{array}\right).
\end{equation}
By Stueckelberg's rule,
\begin{equation}
    \Tilde{V_2V_1}=\tilde{V_2} \tilde{V_1}.
\end{equation}
Again, for any matrix multiplication, the product of two extended real-number matrices is equal to the extended real-number matrix obtained by multiplying the two original matrices, e.g., $\tilde{V_2} \tilde{V_1}=\Tilde{V_2V_1} \xrightarrow{C} V_2V_1$ , where $\xrightarrow{C}$ donate the map from extended matrix to original complex matrix.

Due to the generality of the above analysis, the $V_i$ operation can represent the quantum operation of a single quantum system. The results above indicate that, for a single quantum system, the real-number description of quantum operations (including addition and multiplication) can be transformed from the original complex operation matrices. Alternatively, it can also be obtained by performing the corresponding addition and multiplication on each extended real-number matrix. These both methods are equivalent for a single quantum system.

This consistency is crucial. We will subsequently analyze that, under the assumption of locality (multiple independent systems/sources), the two aforementioned methods of real-number transformation (addition and multiplication) for composite systems described by tensor products are not equivalent generally since 
\begin{equation}
      \text{Map}_R{(V_1\otimes V_1)} =  \tilde{(V_1\otimes V_1)} \ne \tilde{V_1}\otimes \tilde{V_1}. \
\end{equation}
This leads to the results of why complex numbers are necessary in \cite{Renou2021}. To ensure that the real-number description is compatible, one may need to redefine the tensor product for the real-number description \cite{Hita,Hoffreumon2025}. However, the mathematical definition introduces a nonlocal effect in nature. We will analyze this below.

\subsection{ Evolution and expectation value of observables of real-valued quantum theory}\label{s2}

For a single quantum system, the map of evolved quantum state $\rho$ 
 between complex-valued Hilber state and real-valued Hilber space is 
\begin{equation}
    U \rho U^\dagger \xrightarrow{R} \text{Map}_R(U \rho U^\dagger)=  \tilde{U} \tilde{\rho}  \tilde{U^\dagger}.
\end{equation}
For quantum channels, one has 
\begin{equation}
 \mathcal{E}(\rho)=\sum_j K_j \rho K^\dagger_j   \xrightarrow{R} \text{Map}_R(\sum_j K_j \rho K^\dagger_j )=\sum_j \tilde{K_j} \tilde{\rho } \tilde{K}^\dagger_j =:  \mathcal{\tilde{E}}(\tilde{\rho}) .
\end{equation}

Suppose $O$ is Hermitian,  $O=\sum_i o_i \ket{o_i}\bra{o_i}$ where $o_i$ is real number, $O=O_R$. Clearly, expanding $O$ in its eigenbasis, one has
\begin{equation}
    \tilde{O}= \left( \begin{array}{cc}
        O_R &  0 \\
        0 & O_R
    \end{array}\right),
\end{equation}
which is Hermitian. 
In this case, for any $\rho$,
$\Tr(\rho O)=\Tr{(\rho_R+i \rho_I)O}=\Tr(\rho_R O)$, where $\Tr(\rho_I O)=0$. This is because  $\rho_I$ is traceless.
To map the expectation value of observable $O$, one has 
\begin{equation}
\Tr( \mathcal{E}(\rho)O) = \Tr( \mathcal{\tilde{E}}(\tilde{\rho})\tilde{O}). 
\end{equation}
For any density matrix and any observable of a single quantum system, the above equation implies that complex-number quantum theory and real-number quantum theory are consistent.

\section{Composite systems with assumption of Independent Sources  }

\subsection{The entanglement process between independent sources}
So far, we have reproduced the state space axiom (\autoref{s1}), the evolution axiom(\autoref{s2}), the Born Rule (\autoref{s2}), and the Hermitian operator axiom (\autoref{s2}) for the real-valued quantum theory of a single quantum state. The axiom of the tensor product for composite systems does not hold in the real-number Hilbert space mapping, which consequently leads to the inability of real-number quantum theory to describe phenomena such as bi-locality under the assumption of independent sources \cite{Renou2021}.

Suppose $\tilde{\rho_1}$ denote  a  quantum system 1 at position $A$ and $\tilde{\rho_2}$ denote a quantum system 2 at position $B$, where $A$ and $B$ are spacelike separated. If the real-number density matrix $\tilde{\rho_1}$ 
 is the physical description of system 1 and the real-number  density matrix $\tilde{\rho_2}$ 
 is the physical description of system 2, i.e., $\tilde{\rho_1}$ 
 and $\tilde{\rho_2}$ 
 are two independent sources satisfying locality. 
 
A trivial case is that systems 1 and 2 never interact, and we perform only local operations and measurements, real-number quantum theory remains consistent with complex-number quantum theory. Consequently, as in the analysis of a single system, we can reproduce the states, evolution, and measurement outcomes of complex-number quantum theory via an equivalent mapping. That is, the following formula holds

\begin{equation} \label{predict}
    \Tr( \mathcal{E}_1(\rho_1)\otimes \mathcal{E}_2(\rho_2) O_1\otimes O_2) = \Tr( \mathcal{\tilde{E}}_1(\tilde{\rho_1})\otimes\mathcal{\tilde{E}}_2(\tilde{\rho_2})  \tilde{O}_1 \otimes \tilde{O}_2 ). 
\end{equation}
This result can be generalized to $N$-party systems.

 If the tensor product could be extended to real quantum theory in a general case (involving entangled  operations between two and more independent sources), then it should be 
 $\tilde{\rho_1}\otimes \tilde{\rho_2} = \text{Map}_C(\rho_1\otimes \rho_2)=\tilde{(\rho_1\otimes \rho_2) }$.
However, it can be verified that $\text{Map}_R(\rho_1\otimes \rho_2) \ne   \tilde{\rho_1 }\otimes    \tilde{\rho_2 }$, and generally

\begin{equation} \label{entanglementcase}
    \Tr( \mathcal{E}(\rho_{12}) O_1\otimes O_2)\ne \Tr( \mathcal{\tilde{E}}(\tilde{\rho_{12}}) \tilde{O}_1 \otimes \tilde{O}_2 ). 
\end{equation}
This means that, in such a case, we cannot revert it to the description of standard complex-number quantum theory directly.

For general matrix $V_1$ and $V_2$, $V_1\otimes V_2=(V^{(1)}_R+iV^{(1)}_I)\otimes (V^{(2)}_R+iV^{(2)}_I)=V^{(1)}_R\otimes V^{(2)}_R-V^{(1)}_I\otimes V^{(2)}_I+i(V^{(1)}_I \otimes V^{(2)}_R+V^{(1)}_R \otimes V^{(2)}_I)$, resulting
\begin{equation}
  \text{Map}_R(V_1\otimes V_2)=   \left( \begin{array}{cc}
      V^{(1)}_R\otimes V^{(2)}_R-V^{(1)}_I\otimes V^{(2)}_I \quad & \quad
      -V^{(1)}_I \otimes V^{(2)}_R-V^{(1)}_R \otimes V^{(2)}_I\\
       V^{(1)}_I \otimes V^{(2)}_R+V^{(1)}_R \otimes V^{(2)}_I \quad & \quad V^{(1)}_R\otimes V^{(2)}_R-V^{(1)}_I\otimes V^{(2)}_I
    \end{array}\right).
\end{equation}
In contrast, we have 
\begin{equation}
 \tilde{V_1 }\otimes    \tilde{V_2 }=   \left( \begin{array}{cc}
        V^{(1)}_R &  -V^{(1)}_I \\
        V^{(1)}_I & V^{(1)}_R
    \end{array}\right) \otimes \left( \begin{array}{cc}
        V^{(2)}_R &  -V^{(2)}_I \\
        V^{(2)}_I & V^{(2)}_R
    \end{array}\right) = \left( \begin{array}{cc}
        V^{(1)}_R\left( \begin{array}{cc}
        V^{(2)}_R &  -V^{(2)}_I \\
        V^{(2)}_I & V^{(2)}_R
    \end{array}\right) &  V^{(1)}_I \left( \begin{array}{cc}
        V^{(2)}_R &  -V^{(2)}_I \\
        V^{(2)}_I & V^{(2)}_R
    \end{array}\right) \\
        V^{(1)}_I\left( \begin{array}{cc}
        V^{(2)}_R &  -V^{(2)}_I \\
        V^{(2)}_I & V^{(2)}_R
    \end{array}\right) & V^{(1)}_R\left( \begin{array}{cc}
        V^{(2)}_R &  -V^{(2)}_I \\
        V^{(2)}_I & V^{(2)}_R
    \end{array}\right)
    \end{array}\right).
\end{equation}
Obviously, the dimensions of $\text{Map}_RV_1\otimes V_2)$ and $ \tilde{V_1 }\otimes    \tilde{V_2 }$ are inconsistent. The dimensions are not actually the most important thing; what is important is that the framework of real-number quantum theory needs to predict the same properties as complex quantum theory.


Therefore, in general, real-number quantum theory based on the standard tensor product cannot simulate standard complex-number quantum theory when involving entanglement between two independent sources.  A new framework needs to be further analysed.
As demonstrated by \cite{Renou2021}, even with the introduction of infinite-dimensional real-number operations to measure (non-local) observables, there remains a gap between its predictions for the expectation values of actual observables and those of complex-number quantum theory.

\subsection{Modified tensor product in quantum theory of real numbers is a nonlocal map}

As defined in Section II above, the operations in real-number quantum theory should be carefully considered. A simple modified tensor product in real-valued quantum theory can be introduced so that it can be compatible with the predictions of complex-number quantum theory. However, as we show later, the modified tensor product is a nonlocal map.

Suppose there are two pure states $\ket{\psi}$ and $\ket{\phi}$, and they are prepared independently.   In standard quantum theory of complex numbers,  $ \ket{\psi}\otimes \ket{\phi}=\sum_{i=0}^{d-1}(a_i+ib_i)\ket{i} \otimes \sum_{j=0}^{d-1}(c_j+id_j)\ket{j}$ and its real-number state is given as
\begin{equation}
\begin{split}
        \text{Map}_R(\ket{\psi}\otimes \ket{\phi})
        &=\sum_{i=0}^{d-1}\sum_{j=0}^{d-1}(a_ic_j -b_id_j) \ket{0}\ket{ij}+(b_ic_j +a_id_j)\ket{1}\ket{ij}.
\end{split}
\end{equation}
On the other hand,
for the tensor product of the real-number state $\tilde{\ket{\psi}}$ and $\tilde{\ket{\phi}}$, one has

\begin{equation}
\begin{split}
        \tilde{\ket{\psi}}\otimes \tilde{\ket{\phi}}&=\sum_{i=0}^{d-1}a_i\ket{0}\ket{i}+b_i\ket{1}\ket{i} \otimes \sum_{j=0}^{d-1}c_j\ket{0}\ket{j}+d_j\ket{1}\ket{j}\\
        &=\sum_{i=0}^{d-1} \sum_{j=0}^{d-1} (a_ic_j \ket{00} \ket{ij}+b_id_j \ket{11} \ket{ij}) + (a_id_j \ket{01} \ket{ij}+b_ic_j \ket{10} \ket{ij}).
\end{split}
\end{equation}
Compared to $\text{Map}_R(\ket{\psi}\otimes \ket{\phi})$ and $\tilde{\ket{\psi}}\otimes \tilde{\ket{\phi}}$, one can find a nonlinear map $\tilde{\mathcal{P}}$ to make them consistent. 

Specifically,  $\tilde{\mathcal{P}}$ is jointly implemented on the ancillary qubit for both systems, which is defined as  
\begin{equation}
\begin{split}
        &\tilde{\mathcal{P}} (\ket{00})=\ket{0} ;   \tilde{\mathcal{P}} (\ket{11})=-\ket{0}; (even)\\
        &\tilde{\mathcal{P}} (\ket{01})= \tilde{\mathcal{P}} (\ket{10})=\ket{1}({\text{odd}}).
\end{split}
\end{equation}
Obviously, $ \tilde{\mathcal{P}} $ can be considered a parity function to calculate the parity of the ancillary state. On the other hand, $\tilde{\mathcal{P}}$ can be equivalently understood as introducing an ancillary system to entangle with the ancillary qubits of the original systems 1 and 2, and then disentangle itself.
We note that this approach is similar to that in \cite{Hita}, where the authors attempt to localize the process. However, from a physical ontology perspective \cite{Nicholas}, their framework is not entirely local since they introduce a nonlocal map of operation. We discuss this in the following.

A similar definition holds for the bra vector.
Now we have
\begin{equation}
\begin{split}
        \tilde{\mathcal{P}} (\tilde{\ket{\psi}}\otimes \tilde{\ket{\phi}})&= \sum_{i=0}^{d-1} \sum_{j=0}^{d-1} (a_ic_j \tilde{\mathcal{P}} (\ket{00}) \ket{ij}+b_id_j \tilde{\mathcal{P}} (\ket{11}) \ket{ij}) + (a_id_j \tilde{\mathcal{P}} (\ket{01}) \ket{ij}+b_ic_j \tilde{\mathcal{P}} (\ket{10}) \ket{ij})\\
        &=\sum_{i=0}^{d-1} \sum_{j=0}^{d-1} (a_ic_j -b_id_j) \ket{0} \ket{ij} + (a_id_j +b_ic_j )\ket{1} \ket{ij}= \text{Map}_R(\ket{\psi}\otimes \ket{\phi}).
\end{split}
\end{equation}

Since $  \text{Map}_C\tilde{\mathcal{P}} (\tilde{\ket{\psi}}\otimes \tilde{\ket{\phi}})=\ket{\psi}\otimes \ket{\phi}.$ The real number matrix state of ${\tilde{\mathcal{P}} (\tilde{\ket{\psi}}\otimes \tilde{\ket{\phi}})}$ is given as 
\begin{equation}\begin{split}
    \tilde{\rho}[{\tilde{\mathcal{P}} (\tilde{\ket{\psi}}\otimes \tilde{\ket{\phi}})}]=  \text{Map}_R \left(\text{Map}_C({\tilde{\mathcal{P}} (\tilde{\ket{\psi}}\otimes \tilde{\ket{\phi}})}) \text{Map}_C({\tilde{\mathcal{P}} (\tilde{\bra{\psi}}\otimes \tilde{\bra{\phi}})}) \right)=  \text{Map}_R (\rho_1 \otimes \rho_2 ).
    \end{split}
\end{equation}
where $\rho_1=\ket{\psi}\bra{\psi}$ and $\rho_2=\ket{\phi}\bra{\phi}$.
Now, one may introduce a modified tensor product  $\tilde{\otimes}$ for real-number matrix state,  e.g.
\begin{equation}
  \tilde{\rho_1} \tilde{\otimes }\tilde{\rho_2}  :=\tilde{\rho}[{\tilde{\mathcal{P}} (\tilde{\ket{\psi}}\otimes \tilde{\ket{\phi}})}]=\text{Map}_R (\rho_1 \otimes \rho_2).
\end{equation}
Since we have already demonstrated that the addition and multiplication of real-number description on a single system are self-consistent, and now the modified tensor product for pure states is also self-consistent, this modified tensor product can, therefore, be applied to any density matrix (i.e., $\sum_i p_i \rho^{(i)}_A\otimes \rho^{(i)}_{B}$) and other matrices. That is, 

\begin{equation}
     \tilde{V_1 }\tilde{\otimes   } \tilde{V_2 } = \text{Map}_R(V_1\otimes V_2)=  \left( \begin{array}{cc}
      V^{(1)}_R\otimes V^{(2)}_R-V^{(1)}_I\otimes V^{(2)}_I \quad & \quad
      -V^{(1)}_I \otimes V^{(2)}_R-V^{(1)}_R \otimes V^{(2)}_I\\
       V^{(1)}_I \otimes V^{(2)}_R+V^{(1)}_R \otimes V^{(2)}_I \quad & \quad V^{(1)}_R\otimes V^{(2)}_R-V^{(1)}_I\otimes V^{(2)}_I
    \end{array}\right).
\end{equation}

Since $\text{Map}_R(V_1\otimes V_2)=I \otimes \text{Re}(V_1\otimes V_2) +XZ \otimes \text{Im}(V_1\otimes V_2) $, one has

\begin{equation}
   \tilde{V_1 }\tilde{\otimes   } \tilde{V_2 } =I \otimes \text{Re}(V_1\otimes V_2) +XZ \otimes \text{Im}(V_1\otimes V_2) ,
\end{equation}
which is exactly consistent with the combination rule proposed in \cite{Hoffreumon2025}. 

The modified tensor product seems to satisfy the assumption of independent sources. However, it is a fundamental nonlocal map.  
Suppose Alice independently prepares a pair of entangled particles, $a_1$ and $a_2$. Alice sends one of the entangled particles, $a_2$, to Bob, who then entangles it with his particle, $b$. The new tensor rules imply that all operations on the particles are associated with a (ancillary system) subspace. Therefore, even if $a_1$ is very far from $a_2$ and $b$, when a local operation is performed on $a_1$, its operation can not be described locally, and it must correlate with the operations on $a_2$ and $b$ through a hidden nonlocal degree of freedom \cite{Hoffreumon2025}. This degree of freedom may consist of a many-body hidden ancillary system that forms an effective two-dimensional subspace with non-local effects (as shown in \cite{Hita}). This explains proposed real-number quantum theory may not be considered as completely satisfying locality and independent source conditions since both of these papers utilize this nonlocal map for operations.
In a sense, this modified tensor product rule is equivalent to the original Stueckelberg's rule for composite systems with tensor product, where both are non-local.

\section{Discussions and conclusions}
Our result suggests that one can define a new mathematical map so that the quantum theory of real numbers satisfies the independent source hypothesis. 
However, this map is nonlocal in nature \cite{Hita,Hoffreumon2025}, which is somehow equivalent to the original Stueckelberg's rule for composite systems with tensor product.
It is clear that the work of Renou et al. \cite{Renou2021} and the demonstration experiments \cite{Li2022,Chen2022,Wu2022} exclude standard Stueckelberg-type rules of real-number quantum theory, but not the other rules. Just as the Bell test excludes local hidden variable models, it cannot exclude non-local hidden variable models.
Nevertheless, these real-number formulations inevitably introduce hidden nonlocal operations when describing tensor products. Although the assumption of independent sources is generally considered valid, the presence of nonlocal operations appears to contradict this assumption. This, in turn, demonstrates the necessity of complex numbers in quantum theory. Composite systems that involve no entangling interactions can indeed be adequately described by a real-number quantum theory. However, for composite systems (with independent sources) involving entangling operations, the introduction of complex numbers becomes essential. Otherwise, one may concede that a real-number quantum theory would require the incorporation of a hidden nonlocal map to provide an accurate description.
In this sense, complex numbers are seen as indispensable elements of quantum theory.

\section{Acknowledgements}
The authors thank Fulin Wu, Timothee Hoffreumon,  Mischa Woods, Pedro Barrios, Igor Volovich for their comments. This research was made possible
through the generous support of the Gordon and Betty
Moore Foundation, the Eutopia Foundation, and of the
ID 62312 grant from the John Templeton Foundation,
as part of the ‘The Quantum Information Structure of
Spacetime’ Project (QISS). The opinions expressed in
this project/publication are those of the author(s) and
do not necessarily reflect the views of the John Templeton Foundation. 
 C.R. was supported by the National Natural Science Foundation of China (Grants No. 12075245, 12421005 and No. 12247105).

\end{document}